\documentclass[11pt,a4paper]{article}
\usepackage[utf8]{inputenc}
\usepackage{amsmath}
\usepackage{color}
\usepackage{amsfonts}
\usepackage{amssymb}
\usepackage{hyperref}
\usepackage{varwidth}
\usepackage{fancyvrb}
\usepackage{multirow}
\usepackage{graphicx}
\usepackage{caption}
\usepackage{subcaption}
\usepackage{rotating}
\usepackage[noblocks]{authblk}
\newtheorem{theorem}{Theorem}
\newcommand{\R}{\mathbb{R}}
\newcommand{\C}{\mathbb{C}}
\usepackage[left=2cm,right=2cm,top=2cm,bottom=2cm]{geometry}
\usepackage[
natbib=true,maxbibnames=9,
sorting=none,url=false,date=year,doi=false,firstinits=true,isbn=false,bibwarn=true]{biblatex}
\renewbibmacro{in:}{}

\date{}

\newcommand{\mhs}{multivariable hypergeometric series }
\newcommand{\mhf}{multivariable hypergeometric functions }
\newcommand{\mh}{multivariable hypergeometric }

\newcommand{\acs}{analytic continuations }
\newcommand{\g}[1]{\Gamma(#1)}
\newcommand{\p}[2]{(#1)_{#2}}

\begin{document}
	
	\begin{titlepage}

		\begin{center}
			
			\begin{flushright}
				
			\end{flushright}
			
			\begin{flushright}
				
			\end{flushright}
			
			\begin{flushright}
				
			\end{flushright}

 {\Large\bf \texttt{Olsson.wl} : a \textit{Mathematica} package for the computation of linear\\}

 \medskip

  {\Large\bf transformations of multivariable hypergeometric functions \\}

\vspace{1.5cm}

{\bf B. Ananthanarayan$^{a\ast}$, Souvik Bera$^{a\dagger}$, S. Friot$^{b,c\ddagger}$ and Tanay Pathak$^{a\star}$}\\[1.5cm]
{\small $^a$ Centre for High Energy Physics, Indian Institute of Science, \\
	Bangalore-560012, Karnataka, India}\\[0.5cm]
{\small $^b$ Universit\'e Paris-Saclay, CNRS/IN2P3, IJCLab, 91405 Orsay, France } \\[0.5cm]
{\small $^c$ Univ Lyon, Univ Claude Bernard Lyon 1, CNRS/IN2P3, \\
	IP2I Lyon, UMR 5822, F-69622, Villeurbanne, France}
\\[0.5cm]

\end{center}

\begin{abstract}

		 We present the \texttt{Olsson.wl} \textit{Mathematica} package which aims to find linear transformations for some classes of multivariable hypergeometric functions. It is based on a well-known method developed by P. O. M. Olsson in \textbf{J. Math. Phys.} 5, 420 (1964) in order to derive the \acs of the Appell $F_1$ double hypergeometric series from the linear transformations of the Gauss $_2F_1$ hypergeometric function. We provide a brief description of Olsson's method and demonstrate the commands of the package, along with examples. We also provide a companion package, called \texttt{ROC2.wl} and  dedicated to the derivation of the regions of convergence of double hypergeometric series. This package can be used independently of \texttt{Olsson.wl}.
	 \end{abstract}

\vspace{4cm}

\vspace{3cm}
\small{$\ast$ anant@iisc.ac.in }

\small{$\dagger$ souvikbera@iisc.ac.in}

\small{$\ddagger$ samuel.friot@universite-paris-saclay.fr}

\small{$\star$ tanaypathak@iisc.ac.in}

\end{titlepage}

\section{Introduction}

The theory of hypergeometric functions is a classical subject of mathematics which has been extensively studied, but the problem of finding the analytic continuations of multivariable hypergeometric series is still challenging. Even in the simplest one variable case of the Gauss $_2F_1(a,b,c;z)$ hypergeometric series,  which was first introduced by Euler and studied by Gauss more than two hundreds years ago, this nontrivial problem has been fully solved only twenty years ago  \cite{BECKEN2000449}. In the latter work, rapidly converging series representations of the Gauss hypergeometric function were given for arbitrary values of $z$ and all possible constellations of its parameters $a, b$ and $c$ in order to obtain its fast and accurate computation. 

The importance of the Gauss $_2F_1$ function cannot be overstated, but there are many of its multivariable extensions that play an important role as well, not only in mathematics but also in physics where it is not exaggerated to say that hypergeometric functions are ubiquitous. 
It is therefore interesting to focus on these functions and try to get efficient ways to evaluate them. As an example, multivariable hypergeometric functions arise in the context of perturbative calculations in quantum field theory. It seems that Regge was the first to show that Feynman integrals are linked to \mhf \cite{Regge} (for a recent review of the relation between hypergeometric functions and Feynman integrals, we refer the reader to \cite{Kalmykov:2020cqz}). 
Nowadays, methods like the Mellin-Barnes (MB) representation \cite{Smirnov:2012gma,AnantPRL2021}, the method of brackets \cite{GONZALEZ201050,
gonzalez2011generalized,Ananthanarayan:2021not}, the negative-dimension approach \cite{HALLIDAY1987241,suzuki2014massless}, etc. can provide hypergeometric representations of Feynman integrals.  Recently the realization of Feynman integrals as GKZ hypergeometric systems caught physicists attention \cite{delaCruz:2019skx,Klausen:2019hrg,Feng:2019bdx,Tellander:2021xdz}. In all these studies, in order to get numerical values at physical points it may be necessary to perform analytic continuations of the obtained hypergeometric representations. This is why, in addition to the interest in extracting hitherto unknown formulas, in the present work we are mainly concerned by the derivation of linear transformations which can provide \acs of \mh series. Many results can be found in standard references such as  \cite{Bateman:1953,Slater:1966,Exton:1976, Srivastava:1985}. However, as suggested by several recent works (see e.g. \cite{Bezrodnykh:Horn3var,Bezrodnykh:Lauricellafunctions,Bezrodnykh:LauricellaFD,Bezrodnykh:LauricellaFDz_z,Bezrodnykh:Horn_arbitrary_var}), it is obvious that one can still enlarge the list of these formulas. To ease the derivation of the latter is one of the aims of the \textit{Mathematica} package \texttt{Olsson.wl} that we have built and whose description will be shown in detail in the following. The name of this package finds its roots in a method used by Olsson \cite{Olsson64} to find \acs of the Appell $F_1$ double hypergeometric series. Erdélyi pointed out that the transformation theory of \mhf  can be obtained by using the known transformations of hypergeometric functions with a less number of variables when the former can be written as infinite sums of the latter \cite{erdelyi_1948}. In  \cite{Olsson64}, Olsson used the known linear transformations of the Gauss $_2F_1$ function to derive the \acs of the Appell $F_1$ series \cite{Olsson64}. This approach was later used to the study of the three variables extension of $F_1$, namely the Lauricella  $F_D^{(3)}$, and even of its $n$-variable extension $F_D^{(n)}$ by Exton \cite{Exton:1976}. In a similar way, Exton obtained the \acs of the Appell $F_4$ series in \cite{Exton_1995} while the authors of the present paper, in collaboration with Marichev, have considered Olsson's approach in detail on the case of the Appell $F_2$ \cite{Ananthanarayan:2021bqz}. 
Olsson also studied $F_2$ in \cite{Olsson77}, although in the latter work the $F_2$ function is expressed as a Laplace transform of the product of two $_1F_1$ functions (and thus, the linear transformations of $_1F_1$ are used to find the \acs of $F_2$). Olsson's method can be easily applied to a large number of multiple hypergeometric series.

Motivated by the considerations above,
the \texttt{Olsson.wl} package is, to our knowledge, the first attempt to automatize this general approach. It is capable of finding \acs of hypergeometric series of any number of variables, provided that the Gauss $_2F_1$ or the $_{p+1}F_p$ functions can be found as subseries of the latter. In its first version, it deals with generic values of the hypergeometric parameters (logarithmic cases are not considered). It is also possible, by a conjoint use of the companion package \texttt{ROC2.wl} that we also provide in this paper, to evaluate the regions of convergence of the new series obtained from \texttt{Olsson.wl}. In the present version of the package, the regions of convergence can be obtained for double hypergeometric series only (\texttt{ROC3.wl} which will be able to handle the case of hypergeometric series with three variables, is under development \cite{ASST}).

The article is organized as follows: a brief description of Olsson's method is given in section \ref{olsson_section}, which is followed by a detailed demonstration of the \texttt{Olsson.wl} package's commands and algorithm in section \ref{Olsson package}. We then mention some results found using the package in the next section and present in section \ref{roc} the \texttt{ROC2.wl} package, which is followed by the conclusions.

\section{The Olsson method} \label{olsson_section}

This method uses the known transformation theory of hypergeometric functions of lower variables to find transformations of a given hypergeometric function which can be written as an infinite sum of the former. We recapitulate the method following the paper on Appell $F_1$ by Olsson \cite{Olsson64}. The lower variable hypergeometric function here is the well-known Gauss hypergeometric $_2F_1$ function with one variable, whose series representation is
\begin{align}
_2F_1(a, b ; c ; z)=\sum_{s=0}^{\infty} \frac{(a)_{s}(b)_{s}}{(c)_{s} s !} z^{s} \hspace{1cm} |z|<1
\end{align}
where $(a)_s \doteq \frac{\Gamma(a+s)}{\Gamma(a)}$ defines the usual Pochhammer symbol. 

Well-known linear transformations (analytic continuations around $z=1$ and $z\rightarrow\infty$) of the $_2F_1$ are
\begin{align}
		{ }_{2} F_{1}(a, b, c ; z) &=\frac{\Gamma(c) \Gamma(c-a-b)}{\Gamma(c-a) \Gamma(c-b)}{ }_{2} F_{1}(a, b, a+b-c+1 ; 1-z) \nonumber\\
		&+\frac{\Gamma(c) \Gamma(a+b-c)}{\Gamma(a) \Gamma(b)}(1-z)^{c-a-b}{ }_{2} F_{1}(c-a, c-b, c-a-b+1 ; 1-z) \label{ac_2f1at1}\\
		\end{align}
		and
		\begin{align}
		{ }_{2} F_{1}(a, b, c ; z) &=\frac{\Gamma(c) \Gamma(b-a)}{\Gamma(b) \Gamma(c-a)}(-z)^{-a}{ }_{2} F_{1}\left(a, a-c+1, a-b+1 ; \frac{1}{z}\right) \nonumber\\
		&+\frac{\Gamma(c) \Gamma(a-b)}{\Gamma(a) \Gamma(c-b)}(-z)^{-b}{ }_{2} F_{1}\left(b, b-c+1, b-a+1 ; \frac{1}{z}\right)  \label{ac_2f1atinf}\\
\end{align}
Its Euler transformation is
\begin{align}
	{ }_{2} F_{1}(a, b ; c ; z)=(1-z)^{c-a-b}{ }_{2} F_{1}(c-a, c-b ; c ; z)\label{PET3}
\end{align}
and the two related Pfaff transformations are
\begin{align}
	&{ }_{2} F_{1}(a, b ; c ; z)=(1-z)^{-a}{ }_{2} F_{1}\left(a, c-b ; c ; \frac{z}{z-1}\right) \label{PET1}\\
	&{ }_{2} F_{1}(a, b ; c ; z)=(1-z)^{-b}{ }_{2} F_{1}\left(b, c-a ; c ; \frac{z}{z-1}\right)\label{PET2}
\end{align}

Let us now see how these relations can simply be used to derive analytic continuations of the Appell $F_1$ function. 

The latter is defined as
\begin{align}\label{f1f11}
	F_1 (a,b,b',c,x,y)= \sum_{m,n = 0}^{\infty} \dfrac{\p{a}{m+n} \p{b}{m}\p{b'}{n}}{\p{c}{m+n}} \dfrac{x^m}{m!}\dfrac{y^n}{n!}
\end{align}
with region of convergence : $\text{max}\{|x|,|y|\}<1$.\\
Summing over the summation index $n$, one exhibits the $_2F_1$ series inside the summand:
\begin{align}\label{F1in2F1}
	F_1(a,b,b',c,x,y)= \sum_{m} \dfrac{\p{a}{m}\p{b}{m}}{\p{c}{m}} \dfrac{x^m}{m!} {}_2 F_1 (a+m,b',c+m,y)
\end{align}
Making use of  eq. \eqref{ac_2f1at1} yields the analytic continuation of Appell $F_1$ around $(0,1)$,
\begin{align}\label{F1at0inf}
&F_1= \frac{\Gamma \left(-a+c-b'\right)}{\Gamma (c-a)}\sum_{m=0}^{\infty}	\frac{x^m (a)_m (b)_m \Gamma (c+m)  }{m!  (c)_m \Gamma \left(c+m-b'\right)} \, _2F_1\left(a+m,b';a-c+b'+1;1-y\right)\nonumber\\
	&+ (1-y)^{-a-b'+c} \frac{\Gamma \left(a-c+b'\right)}{\Gamma \left(b'\right)} \sum_{m=0}^{\infty}\frac{x^m (a)_m (b)_m \Gamma (c+m) }{m! \Gamma (a+m)  (c)_m}  \, _2F_1\left(c-a,c+m-b';-a+c-b'+1;1-y\right)
\end{align}
which simplifies to
\begin{align}\label{F1at0infsim}
	F_1(a,b,b',c,x,y)=	&(1-y)^{-a-b'+c} \frac{\Gamma (c)  \Gamma \left(a-c+b'\right)}{\Gamma (a) \Gamma \left(b'\right)} \sum_{m,n = 0}^{\infty} \frac{(b)_m (c-a)_n \left(c-b'\right)_{m+n}}{ \left(c-b'\right)_m \left(-a+c-b'+1\right)_n} \frac{x^m (1-y)^n }{m! n!} \nonumber\\
	&+ \frac{\Gamma (c) \Gamma \left(-a+c-b'\right)}{\Gamma (c-a) \Gamma \left(c-b'\right)} \sum_{m,n = 0}^{\infty} \frac{ (b)_m (a)_{m+n} \left(b'\right)_n}{ \left(c-b'\right)_m \left(a-c+b'+1\right)_n} \frac{x^m (1-y)^n}{m! n!}
\end{align}
This is the same expression as Eq. (9) of \cite{Olsson64}, valid for $|x|+|1-y|<1$. Both the series in the right hand side are Appell $F_2$ functions, which in turn can be expressed in terms of Appell $F_1$ and Horn $G_2$ series. 
In order to get \acs around other singular points, one proceeds in a similar fashion. Using Eqs. \eqref{F1in2F1} and \eqref{ac_2f1atinf} the analytic continuation of $F_1$ around $(0,\infty)$ can be found. Further applying Eq. \eqref{ac_2f1atinf} on Eq. \eqref{F1at0inf} another analytic continuation around $(0,\infty)$ can be derived, etc. We do not dig into the \acs of $F_1$ any further. The reader is referred to \cite{Olsson64} for more details.

The pattern is clear. For the hypergeometric series $F_1$ with two variables, the summation over one index is taken, resulting in the appearance of the Gauss $_2F_1$ inside the summand. Then a good use of the transformation theory of $_2F_1$ yields the \acs of the hypergeometric series we started with. This is a very general procedure to obtain \acs of \mh series. One will note that it is not necessary that $_2F_1$ appears inside the summand. It could be any other hypergeometric series whose transformation theory is known. However, in the present version of the \texttt{Olsson.wl} package, to be described in the next section, we focus on the cases where $_2F_1$ and $_{p}F_{p-1}$ appear as subseries.

\section{The \texttt{Olsson.wl} \textit{Mathematica} package \label{Olsson package}}

In this section, we present the \texttt{Olsson.wl} package by giving a brief description of its commands and their working procedure with illustrated examples. The package can work on \textit{Mathematica} 11.3 or higher version.

After downloading the package, the path is set up for \texttt{Olsson.wl} as,

\begin{Verbatim}[fontsize=\small,frame=single]
In[]:= <<"MyDirectory/Olsson.wl"
\end{Verbatim}

We start by describing some well-known properties and formulae of the Euler gamma function and Pochhammer symbol that play a key role in handling hypergeometric functions. These are implemented in \textit{Mathematica} as substitution rules and  are used repeatedly in this package. 

\subsection{Substitution rules}\label{substitutionrules}
In what follows we use \texttt{m,n} and \texttt{p} as summation indices, each of them taking non negative integer values.

\subsubsection{\textbf{\texttt{gammatoPoch}  }}
Gamma functions appearing in the calculations are transformed into Pochhammer symbols using the \texttt{gammatoPoch} command. 

This command generates the two substitution rules 
\begin{align}\label{gammatoPoch}
	\Gamma(a+n)\rightarrow\p{a}{n} \Gamma(a), \hspace{2cm} \p{a+n}{m}\rightarrow\frac{(a)_{m+n}}{(a)_n}
\end{align}
The argument of \texttt{gammatoPoch} is a list and it generates such transformation rules for each entry of its argument list. For example,
\begin{Verbatim}[fontsize=\small,frame=single]
In[]:=Gamma[a + n + 2 m] /. gammatoPoch[{n}][[1]]

Out[]=Gamma[a + 2 m] Pochhammer[a + 2 m, n]

In[]:=Pochhammer[a + n, m] /. gammatoPoch[{n}][[2]]

Out[]=Pochhammer[a, m + n]/Pochhammer[a, n]

In[]:=Gamma[a + n + 2 m] /. gammatoPoch[{n, m}]

Out[]=Gamma[a + 2 m] Pochhammer[a + 2 m, n]

In[]:=Gamma[a + n + 2 m] //. gammatoPoch[{n, m}]

Out[]=Gamma[a] Pochhammer[a, 2 m + n]
\end{Verbatim}
Note the repetitive use of \texttt{gammatoPoch[\{n, m\}]} in the last input.

\subsubsection{\textbf{\texttt{positivePoch} }}
The substitution rule,  
\begin{align}
	\p{a}{-n}\rightarrow \frac{(-1)^n}{\p{1-a}{n}}, \hspace{1cm} a\notin \mathbb{Z} ~\text{and}~ n\in \mathbb{N}
\end{align}
is implemented in the \texttt{positivePoch} command. It also takes a list as argument and generates the above substitution for each of the entries of that list.
\begin{Verbatim}[fontsize=\small,frame=single]
In[]:=Pochhammer[a + b, -m - 2 n] /. positivepoch[{m, n}]

Out[]=(-1)^(m + 2 n)/Pochhammer[1 - a - b, m + 2 n]
\end{Verbatim}

\subsubsection{\textbf{\texttt{Pochdim}  }}

The Pochhammer dimidiation  formula,
 \begin{align}
	\p{a}{2 m} \rightarrow 2^{2 m} \left(\frac{a}{2}+\frac{1}{2}\right)_m \left(\frac{a}{2}\right)_m
\end{align}
is implemented in the \texttt{Pochdim} command.
For example,

\begin{Verbatim}[fontsize = \small,frame=single]
In[]:=Pochhammer[a, 3 m] /. Pochdim

Out[]=3^(3 m) Pochhammer[1/3 + a/3, m] Pochhammer[2/3 + a/3, m] Pochhammer[a/3, m]

In[]:=Pochhammer[a, 3 n + 9 m] /. Pochdim

Out[]= Pochhammer[a, 9 m + 3 n]

In[]:=Simplify /@ (Pochhammer[a, 3 n + 9 m]) /. Pochdim

Out[]=3^(3 (3 m + n))
Pochhammer[1/3 + a/3, 3 m + n] Pochhammer[2/3 + a/3, 3 m + n] Pochhammer[a/3, 3 m + n]

In[]:=Pochhammer[a, -3 m] /. Pochdim

Out[]=Pochhammer[a, -3 m]
\end{Verbatim}
Note that the \texttt{Simplify} command simplifies the argument of the Pochhammer parameter above to \texttt{Pochhammer[a,3 (3 m + n)]}, which matches the pattern of \texttt{Pochdim}.

\subsubsection{\textbf{\texttt{unsim} }}
This command does the opposite of the second expression of Eq. \eqref{gammatoPoch}:
\begin{align}
	(a)_{m+n} \rightarrow (a)_n (a+n)_m
\end{align}
It only takes one summation index as argument.
\begin{Verbatim}[fontsize = \small,frame=single]
In[]:= Pochhammer[a, 2 m + 3 n + p] /. unsim[n]

Out[]=Pochhammer[a, 2 m + p] Pochhammer[a + 2 m + p, 3 n]
\end{Verbatim}

\subsection{Some other commands}

Here we mention some secondary commands of this package. 
\subsubsection{\textbf{\texttt{characteristiclist}}}
It gives the characteristic list of a series. For example,
\begin{Verbatim}[fontsize=\small,frame=single]
In[]:=characteristiclist[{m, p}, (Pochhammer[a, m + p] Pochhammer[b, m] Pochhammer[b1, p] 
x^m y^p)/(Pochhammer[c, m + p] m! p!)]

Out[]={{{{m + p, m, p}, {m + p}}, {x, y}}}
\end{Verbatim}
The first entry of the list given in the output is the characteristic list of the given series, which here is Appell $F_1$, and the second entry is the arguments of the series.

\subsubsection{\texttt{hypgeoinf}}
The \acs of $_pF_{p-1}$ at infinity can be found by this command, for generic values of its parameters. For example,
\begin{Verbatim}[fontsize=\small,frame=single]
In[]:=hypgeoinf[3, 2, z]
	
Out[]= (((-z)^-Subscript[a,1]) Gamma[-Subscript[a,1] 
+ Subscript[a,2]] Gamma[-Subscript[a,1] + 
Subscript[a,3]]Gamma[Subscript[b,1]]Gamma[Subscript[b, 2]]HypergeometricPFQ[...,1/z])/(...)
+...
	
(*output is suppressed due to long expression*)
\end{Verbatim}

The output is the expression of the analytic continuation of $_3F_2[\{a_1,a_2,a_3\},\{b_1,b_2\},z]$ around $z=\infty$.

\subsubsection{\texttt{hypgeoatinf}}
This gives the analytic continuation of any $_pF_{p-1}$  around $z=\infty$ when the parameters and argument are specified. For example,
\begin{Verbatim}[fontsize=\small,frame=single]
In[]:=hypgeoatinf[{a, b, c}, {d, e}, z] 
	
Out[]= ((-z)^-a Gamma[-a + b] Gamma[-a + c] Gamma[d] Gamma[e] 
HypergeometricPFQ[{a, 1 + a - d, 1 + a - e}, {1 + a - b, 1 + a - c}, 1/z])/(...)+...
	
(*output is suppressed*)
\end{Verbatim}

The output is thus the same as the output of \texttt{hypgeoinf[3, 2, z]} after the replacement of \texttt{a$_\texttt{1}$->a}, \texttt{a$_\texttt{2}$->b}, etc.

We are now in a position to show the main command of the package, namely the \texttt{Olsson} command. The above replacement rules and modules are used in the latter.

\subsection{The \texttt{Olsson} command}

The \texttt{Olsson} command can be used to find analytic continuations of some classes of multivariable hypergeometric series of any number of variables, in the neighbourhood of singular points. It is based on the method used by Olsson in \cite{Olsson64}. Therefore, in the first version of this package, the main constraint that the multivariable hypergeometric series must  satisfy, in order to be handled by the package, is that they can be written as sums of Gauss $_2F_1$ or $_pF_{p-1}$ hypergeometric series.

 The \texttt{Olsson} command can be called as follows,
\begin{align*}
\texttt{Olsson[q, summation-indices-list, expression, options]}
\end{align*}
The first argument \texttt{q} is an integer that dictates the summation index of the series with respect to which the analytic continuation will be performed. 

The summation indices in \texttt{summation-indices-list} can take only non negative integer values. As an example, if \texttt{\{m,n\}} is the list of summation indices and \texttt{q} is \texttt{1} (or \texttt{2}), then the analytic continuation is obtained for the series associated with the first (or second) summation index, i.e. \texttt{m}. (or \texttt{n}). Clearly \texttt{q} can take values from \texttt{1} to \texttt{Length[summation-indices-list]}.

The available options of the \texttt{Olsson} command are,
\begin{align*}
	\texttt{sum,one,inf,PET1,PET2,PET3,sim,roc}
\end{align*}

We now demonstrate the working principle of each of these options.  Although the \texttt{Olsson} command works for multivariable hypergeometric series, for the sake of illustration we take here as input the Appell $F_2$ double hypergeometric series, which we have recently considered in detail in \cite{Ananthanarayan:2021bqz}. The default value of each of the options above is \texttt{False}.

Let us begin with the option \texttt{sum} and see what can be obtained from it on $F_2(a, b_1,b_2;c_1,c_2;x,y)$. 

\begin{Verbatim}[fontsize=\small,frame=single]
In[]:=appellF2 =  (Pochhammer[a, m + p] Pochhammer[b1, m] Pochhammer[b2, p] x^m y^p)/
(Pochhammer[c1, m] Pochhammer[c2, p] m! p!);	
	
In[]:=Olsson[1, {m, p}, appellF2, {sum -> True}]
	
Out[]=(y^p HypergeometricPFQ[{b1, a + p}, {c1}, x] Pochhammer[a, p] Pochhammer[b2, p])/
(p! Pochhammer[c2, p])
\end{Verbatim}

In this example, \texttt{sum-> True} takes the summation over \texttt{m} (i.e. the first entry of the summation indices list \texttt{\string{m,p\string}}) and expresses it as \texttt{HypergeometricPFQ} function. Similarly, in order to take the summation over \texttt{p}, one would have to put \texttt{2} as \texttt{q} or alter the order of  the \texttt{summation-indices-list} to \texttt{\string{p,m\string}}, as shown below:

\begin{Verbatim}[fontsize=\small,frame=single]
In[]:=Olsson[2, {m, p},appellF2, {sum -> True}]
	
Out[]=(x^m HypergeometricPFQ[{b2, a + m}, {c2}, y] Pochhammer[a, m] Pochhammer[b1, m])/
(m! Pochhammer[c1, m])
\end{Verbatim}

Let us now consider the option \texttt{one}.

This option can be used to find the analytic continuation around $x=1$ (or $y=1$), by an application of Eq.\eqref{ac_2f1at1}. For this to be performed, one has to select \texttt{one-> True}, which, on our last example, gives

\begin{Verbatim}[fontsize=\small,frame=single]
In[]:=Olsson[2, {m, p}, appellF2, {one -> True}] // Expand

Out[]=(x^m Gamma[c2] Gamma[-a - b2 + c2 - m] HypergeometricPFQ[...,1-y]...)+
(x^m(1-y)^(-a-b2+c2-m)Gamma[c2] Gamma[a + b2 - c2 + m] HypergeometricPFQ[...,1-y]...)
\end{Verbatim}

where we have suppressed part of the output in order to improve the readability. 

In a similar way, if one is interested in the analytic continuation as $x\rightarrow\infty$ (or $y\rightarrow\infty$), one will want to use Eq.\eqref{ac_2f1atinf}, which can be accomplished by selecting the \texttt{inf-> True} option:
\begin{Verbatim}[fontsize=\small,frame=single]
In[]:=Olsson[2, {m, p}, appellF2, {inf -> True}] // Expand
	
Out[]=(x^m (-y)^-b2 Gamma[c2] Gamma[a - b2 + m] HypergeometricPFQ[...,1/y]...) + 
(x^m (-y)^(-a - m) Gamma[c2] Gamma[-a + b2 - m] HypergeometricPFQ[...,1/y]...)
\end{Verbatim}

Note that the above expressions are not simplified, as the summation index \texttt{m} appears inside some gamma functions. All the substitution rules mentioned in section \eqref{substitutionrules} are combined in suitable order to form the option \texttt{sim}. So we use \texttt{sim-> True} to simplify the expressions,

\begin{Verbatim}[fontsize=\small,frame=single]
In[]:=Olsson[2, {m, p}, appellF2, {inf -> True, sim -> True}] // Expand
	
Out[]=(x^m (-y)^(-a-m)y^-p Gamma[-a + b2]Gamma[c2] Pochhammer[a,m + p]Pochhammer[b1,m]
Pochhammer[1 + a - c2, m + p])/(m! p! Gamma[b2] Gamma[-a + c2] Pochhammer[1+a-b2, m + p]
Pochhammer[c1, m]) + ...
\end{Verbatim}

The options \texttt{PET1,PET2,PET3} respectively correspond to the use of Euler and Pfaff transformations given in Eqs.\eqref{PET1}, \eqref{PET2} and \eqref{PET3}. These can be used similarly as the options \texttt{one} or \texttt{inf}, along with the \texttt{sim} option. 

The last option, \texttt{roc}, can be used to find the region of convergence of (only) double hypergeometric series (its extension to hypergeometric series having more than two variables is under consideration). It can be used as \texttt{roc-> True}. The common region of convergence is obtained by this option if several double hypergeometric series appear in the calculations. Note that the package \texttt{ROC2.wl} is needed for this option to work. The package \texttt{ROC2.wl} must be kept in the same path as \texttt{Olsson.wl}. Detailed description of the \texttt{ROC2.wl} can be found below in section \eqref{roc}.

\begin{Verbatim}[fontsize=\small,frame=single]
In[]:=Olsson[2,{m,p},appellF2,{inf->True,sim->True,roc->True}]//Simplify //Expand

Out[]:={Abs[x] < 1 && Abs[x/y] < 1 && 1/Abs[y] < 1 && 1/Abs[y] < 1/(1 + Abs[x]) 
&& Abs[x/y] + 1/Abs[y] < 1, x^m (-y)^(-a - m) y^-p Gamma[-a + b2]...}
\end{Verbatim} 

The second entry of the above output is the analytic continuation of the Appell $F_2$ at $y\rightarrow\infty$. There are two series in the analytic continuation. The first entry  is the common region of convergence of those two series.

Various linear transformations of multivariables hypergeometric functions can thus be found by repetitive use of all these options.

\subsection{\texttt{callroc} \label{callroc}}
Regions of convergence of double hypergeometric series can be found by using the  \texttt{callroc} command as follows:

\begin{Verbatim}[fontsize=\small,frame=single]
In[]:=callroc[{m, p}, appellF2] // FullSimplify
	
Out[]:={Abs[x] + Abs[y] < 1}
\end{Verbatim}

The command \texttt{callroc} can take a sum of the two or more double hypergeometric series as input. In this case, it will give, as an output, the region of convergence of each of the series in a list. The \texttt{ROC2.wl} package is needed for this command to work.

\subsection{ \textbf{\texttt{serrecog}} }
While building transformations of \mh functions, it can be advantageous to automatically identify the functions that appear in the derived expressions. The \texttt{serrecog} command helps to recognize some specific classes of hypergeometric series. Let us see this on the simple example below:

\begin{Verbatim}[fontsize=\small,frame=single]
In[20]:=serrecog[{m, p}, appellF2]

Out[20]={"F2", 1, (x^m y^p Pochhammer[a, m + p] Pochhammer[b, m] Pochhammer[b1, p])/(
m! p! Pochhammer[c, m] Pochhammer[c1, p])}
\end{Verbatim} 
The output is a list containing the name of the series \texttt{"F2"}, its prefactor and the expression of the series. 
This command first separates the prefactors, if any, from the series. The Pochhammer symbols are rearranged using the substitution rules in such a way that the first summation index comes with a positive coefficient in their argument.  Then the characteristic list is found and matched with the pre-stored list of characteristic lists. If any matching is found, it is prompted in the output. Otherwise it gives the characteristic list in the ordered mentioned above. If desired, the characteristic list of the ``unknown" series can be appended to the pre-stored list and the command will recognize it when encountered later.  This is what we explain in the next section.

\subsubsection{Adding new characteristic lists }

The ordered characteristic list of the 14 Appell-Horn series, Kamp\'e de F\'eriet series (with two variables) and four Lauricella series in $N$ variables are stored in the \texttt{Olsson.wl}. However, the user may add the characteristic list of any multivariable hypergeometric series in this list, as we show now with the example of `unknown series', which is not recognized as its characteristic list does not belong to pre-stored list in \texttt{Olsson.wl}:
\begin{Verbatim}[fontsize=\small,frame=single]
In[]:=serrecog[{m, n}, (Pochhammer[a, m + n] Pochhammer[b, 2 m] Pochhammer[c, n] x^m y^n)/
(Pochhammer[d, m + n] m! n! Pochhammer[e, m])]
	
Unknown series! 
{{2 m,n,m+n},{m,m+n}}
	
Out[]={1, (x^m y^n Pochhammer[a, m + n] Pochhammer[b, 2 m] Pochhammer[c, n])/
()m! n! Pochhammer[d, m + n] Pochhammer[e, m])}
\end{Verbatim} 
As the series is not recognized by the \texttt{serrecog} command, a message `Unknown series' is showed along with the ordered characteristic list.  
To add it, one can copy the ordered characteristic list and paste it inside the table for 2 variables (\texttt{table2var}) as 
\begin{align*}
	\texttt{\{\{\{2 m,n,m+n\}, \{m,m+n\}\},"new" (*or any other name*)\}}
\end{align*} 

The package will recognize the series after being reloaded. 

\begin{Verbatim}[fontsize=\small,frame=single]
In[]:=Quit[]
	
In[]:=<<"MypathforOlsson/Olsson.wl"
	
In[]:=serrecog[{m, n}, (Pochhammer[a, m + n] Pochhammer[b, 2 m] Pochhammer[c, n] x^m y^n)/
(Pochhammer[d, m + n] m! n! Pochhammer[e, m])]
	
Out[]={"new", 1, (x^m y^n Pochhammer[a, m + n] Pochhammer[b, 2 m] Pochhammer[c, n])/
(m! n! Pochhammer[d, m + n] Pochhammer[e, m])}
\end{Verbatim} 

The same procedure works for series with three variables by putting the characteristic lists of unknown series inside the table for three variables (\texttt{table3var}). Note that, in order to add new characteristic lists one must use \texttt{m,n} as indices for two variable series and \texttt{m,n,p} as indices for three variable series.  For the present version of \texttt{Olsson.wl}, we restrict the \texttt{serrecog} command to recognize unknown series with variables up to three, with the exception of Lauricella functions, which can be recognized in $N$ variables, as mentioned above.

\subsection{\texttt{serrecog2var}}

There is a special command for the more explicit recognition of two variable series. We demonstrate this below
\begin{Verbatim}[fontsize=\small,frame=single]
In[]:=appellF2 = (Pochhammer[a, m + p] Pochhammer[b1, m] Pochhammer[b2, p] x^m y^p)/
(Pochhammer[c1, m] Pochhammer[c2, p] m! p!);
	
In[]:=serrecog2var[{m, p}, appellF2]
	
Out[]=F2[a, b1, b2, c1, c2, x, y]
\end{Verbatim} 
Note the difference in the outputs of \texttt{serrecog} and \texttt{serrecog2var}. This command can recognize the 14 Appell-Horn series and arbitrary Kamp\'e de F\'eriet series in two variables.

\section{Results obtained using \texttt{Olsson.wl}}

The package is capable of finding new transformations of \mh functions. Thus, after having reproduced some results of the literature for the Appell $F_2$ and $F_4$ functions, we looked at less studied functions, such as the Horn $H_1$ and $H_5$ functions. We also tested our package on the Lauricella functions of 3 variables. The new transformations obtained were numerically checked when possible. It turns out that, using this method, one can find the \acs that cover the whole space of its arguments for $F_2$, $F_4$ and $H_1$ series, except at some specific points. For the case of \acs of $H_5$, more work is needed to cover the whole space of its arguments, since white zones still appear. The analytic continuations of other order two complete double hypergeometric functions can be found by  using their relation to the Appell $F_2$ function, as discussed in \cite{Ananthanarayan:2021bqz}.

\section{Regions of Convergence (ROC) of double hypergeometric series\label{roc}} 
As already briefly described in section \ref{callroc}, the command \texttt{callroc} and the \texttt{roc} option in the \texttt{Olsson} command can be used to find the regions of convergence of double hypergeometric series. The is performed by another package that we have built, \texttt{ROC2.wl}, for which we provide the theoretical backgrounds, algorithm and usage below. We note here that, if needed, this package can be used separately from the \texttt{Olsson.wl} package.

The following theorems are useful in determining the ROC of double hypergeometric series  \cite{Srivastava:1985}.
\vspace{0.5cm}

\textbf{Cancellation of Parameters}

\begin{theorem}\label{copara}
	The region of convergence of a hypergeometric series is independent of its parameters, exceptional parameter values being excluded.
\end{theorem}
For example, let us consider again the Appell $F_2$ series
\begin{equation}
F_{2}(a,b,c;d,e;x,y)= \sum_{m,n=0}^{\infty} \frac{\p{a}{m}\p{b}{n}\p{c}{m+n}}{\p{d}{m} \p{e}{n} m!n!}x^{m}y^{n}
\end{equation}
whose well-known ROC is $ |x| + |y|< 1$.\\\\
If we now consider the following Kampé de Fériet series
\begin{equation}
F^{1:1:2}_{0:1:2}(a_1,b_1,c_1,d_1;a_2,b_2,c_2;x,y)= \sum_{m,n=0}^{\infty} \frac{\p{a_1}{m}\p{b_1}{n}\p{c_1}{m+n}\p{d_1}{n}}{\p{a_2}{m} \p{b_2}{n}\p{c_2}{n} m!n!}x^{m}y^{n}
\end{equation}
we see, from Theorem \ref{copara}, that we can choose $d_1= c_2$, so that in fact both series above have the same ROC.
\vspace{0.5cm}

\textbf{Horn's theorem for double hypergeometric series}
\vspace{0.5cm}

\noindent Now we discuss a general theorem to obtain the ROC of double hypergeometric series.

Consider a double hypergeometric series 
\begin{equation}
F(x,y) = \sum_{m,n=0}^{\infty} C(m,n) x^{m}y^{n}
\end{equation}
and define
\begin{equation}
\begin{aligned}
f(m,n)&\doteq\frac{C(m+1,n)}{C(m,n)}\\
g(m,n)&\doteq\frac{C(m,n+1)}{C(m,n)}
\end{aligned}
\end{equation}
which are rational functions.
Let us further define
\begin{equation}
\rho(m,n)= |\lim_{u \rightarrow \infty}f(mu,nu)|^{-1},\quad m>0, n\geq 0
\end{equation}
\begin{equation}
\sigma(m,n)= |\lim_{u \rightarrow \infty}g(mu,nu)|^{-1},\quad m\geq0, n> 0
\end{equation}
which are rational functions too. From these functions, one constructs the two subsets of $\R_{+}^{2}$
\begin{equation}
\begin{aligned}
C &= \{(r,s)| 0 < r < \rho(1,0) \wedge 0< s<\sigma(0,1)\}\\
&\doteq K[\rho(1,0),\sigma(0,1)]
\end{aligned}
\end{equation}
and
\begin{equation}
Z= \{(r,s)| \forall (m,n) \in \R_{+}^{2}:r<\rho(m,n)\lor 0<s<\sigma(m,n)\}
\end{equation}
The Horn's theorem for double hypergeometric series can now be stated as
\begin{theorem}
	The union of $Z \cap C$ and its projections upon the co-ordinate axes is the representation in the absolute quadrant $\R_{+}^{2}$ of the region of convergence in $\C^{2}$ for the series F.
\end{theorem}

\subsection{The usage of \texttt{ROC2.wl}\label{algo}}

The \texttt{ROC2.wl} package is a separate package that can be called without \texttt{Olsson.wl} as follows:
\begin{Verbatim}[fontsize=\small,frame=single]
In[]:= <<"MypathforROC2/ROC2.wl"
\end{Verbatim}

Let us briefly mention the steps of the algorithm used in the \texttt{ROC2.wl} package.
\begin{enumerate}
	\item Use the cancellation of parameters on the characteristic list of the studied series.
	\item Construct the rational functions.
	\item Find the rectangle $K[R,S]=K[\rho(1,0),\sigma(0,1)]$.
	\item Evaluate the conditions from the rational functions.
	\item Take the absolute value of the rational functions depending on the conditions to find the cases.
	\item Eliminate the summation indices to find the boundaries.
	\item Check the degrees of the variables.
	\begin{itemize}
		\item If the degrees are greater  or equal to  $1$ and identical for both the variables, then solve for $|y|$.
		\item If the degrees are greater or equal to $1$ and not identical for both the variables, then solve for the variable that has the maximum degree.
		\item If the degrees ($\text{deg}$) are less than  $1$, then solve for $|y|^{1/\text{deg}}$.
	\end{itemize}
	
	Note that the boundaries are solved with the constraint that $|x|$ should lie within $R$.
	Find the region  by replacing the equality sign by less than sign.
	
\end{enumerate}


\begin{figure}
	\includegraphics[width=\textwidth]{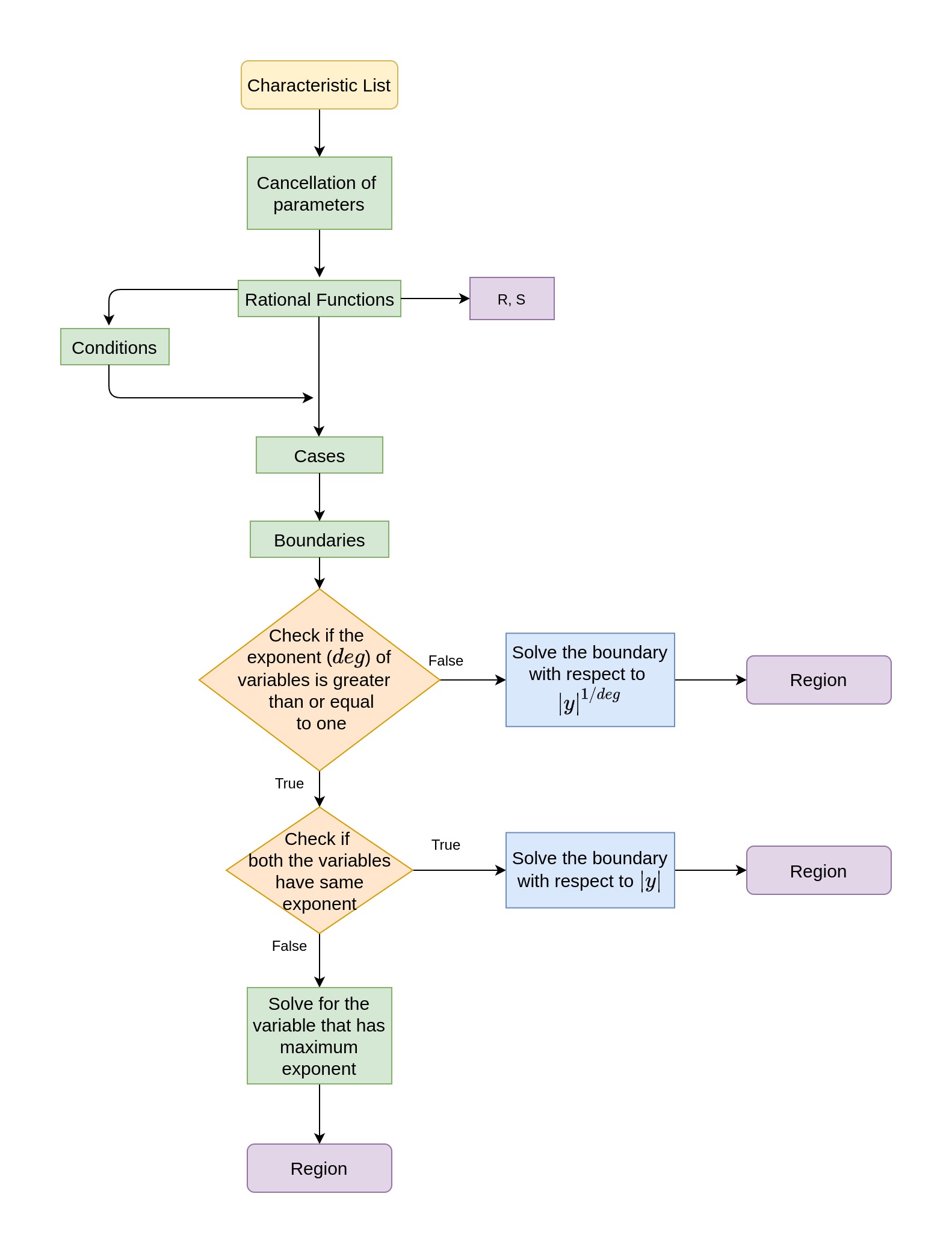}
	\caption{Flow chart of the \texttt{ROC2} algorithm }
	\label{roc2_flowchart}
\end{figure}


The flow chart of the above algorithm is shown in figure \ref{roc2_flowchart}. We now show how in the following section how to call the \texttt{ROC2} command which applies the algorithm above.

\subsubsection{Input}

The algorithm of section \ref{algo} is implemented in the \texttt{ROC2} command which takes the form
\begin{center}
\texttt{ROC2[m,n,num\_List,denom\_List,X,Y]}
\end{center}
where \texttt{m} and \texttt{n} denote the two summation indices and \texttt{X} and \texttt{Y} the two variables of the studied double hypergeometric series. The two lists \texttt{num\_List} and \texttt{denom\_List} split the characteristic list of the series into two sets corresponding to the gamma functions appearing in the numerator and in the denominator of the series. 

For example, for the $F_2$ series, one has
\begin{align*}
F_2 (a,b_1,b_2,c_1,c_2,x,y) = \sum_{m,n=0}^{\infty} \dfrac{\p{a}{m+n}\p{b_1}{m}\p{b_2}{n}}{\p{c_1}{m}\p{c_2}{n}} \dfrac{x^m}{m!}\dfrac{y^n}{n!}
\end{align*}  
with \texttt{num\_List} = $\{m+n,m,n\}$, \texttt{denom\_List} =  $\{m,n\}$, \texttt{X}=$x$ and \texttt{Y}=$y$.

For the Appell $F_3$ series, written in terms of gamma functions,
\begin{align*}
&F_3(a_1,a_2,b_1,b_2,c,x,y) =\\ &\dfrac{\g{c}}{\g{a_1}\g{a_2}\g{b_1}\g{b_2}}\sum_{m,n=0}^{\infty} \dfrac{\g{a_1+m}\g{a_2+m}\g{b_1+n}\g{b_2+n}}{\g{c+m+n}}\dfrac{x^m}{m!}\dfrac{y^n}{n!}
\end{align*}
one has \texttt{num\_List} = $\{m,m,n,n\}$, \texttt{denom\_List} =  $\{m+n\}$, \texttt{X}=$x$ and \texttt{Y}=$y$.

Note that we do not include the $m!$ and $n!$ in \texttt{denom\_List}. The order of elements in \texttt{num\_List} and \texttt{denom\_List} does not matter and if there is no denominator in a series then  \texttt{denom\_List} can be taken as $\{1\}$.

For the above two examples, the call of \texttt{ROC2} is thus simply
\begin{align*}
&\texttt{ROC2[m,n,\{m+n,m,n\},\{m,n\},x,y]} \hspace{1cm}\text{for} ~F_2\\
&\texttt{ROC2[m,n,\{m,m,n,n\},\{m+n\},x,y]} \hspace{1cm}\text{for} ~F_3
\end{align*}

One may encounter series with variables which are functions of both $x$ and $y$. The package can evaluate the ROC for those cases too. For example, the ROC of $F_2 (a,b_1,b_2,c_1,c_2,x/y,y)$ would be obtained from \texttt{ROC2[m,n,\{m+n,m,n\},\{m,n\},x/y,y]}.

\subsubsection{Output}\label{sec_o/p}
The output of the  \texttt{ROC2} command gives the rectangle $K[\rho(1,0),\sigma(0,1)]$, the Cartesian curve which will help finding the boundary of $Z$ and the ROC. If the ROC for a series is the rectangle itself then the Cartesian curve is not given. The ROC can be simplified further using the inbuilt \textit{Mathematica} \texttt{FullSimply} command. Let us show the output for the above mentioned simple examples, before moving to a less trivial case in the next section.

\begin{Verbatim}[fontsize=\small,frame=single]
In[]:=ROC2[m, n, {m, n, m + n}, {m, n}, x, y](*ROC of F2 *)

Out[]={"{R,S}, Cartesian Curve, ROC -> ", {1,1}, {Abs[x] + Abs[y] == 1},
{Abs[x] < 1 && Abs[y] < 1 && Abs[y] < 1 - Abs[x]}}

In[]:=ROC2[m, n, {m, n, m, n}, {m + n}, x, y](*ROC of F3*)

Out[]={"{R,S}, Cartesian Curve, ROC -> ", {1,1}, {}, {Abs[x] < 1 && Abs[y] < 1}}

In[]:=ROC2[m, n, {m, n, m + n}, {m, n}, x/y, y]

Out[]={"{R,S}, Cartesian Curve, ROC -> ", {1, 1}, {Abs[x/y] + Abs[y] == 1},
{Abs[x/y] < 1 && Abs[y] < 1 && Abs[y] < 1 - Abs[x/y]}}

In[]:=FullSimplify[Last[%]](* Further simplication *)

Out[]={Abs[x/y] + Abs[y] < 1}
\end{Verbatim}


\subsubsection{The Horn $H_5$ series}
In this section, we show some details about the different steps of the algorithm, when applied to the non-trivial case of the Horn $H_5$ series.
The latter is defined as
\begin{align*}
H_5(a,b,c,x,y) = \sum_{m,n=0}^{\infty} \dfrac{\p{a}{2m+n}\p{b}{n-m}}{\p{c}{n}} \dfrac{x^m}{m!}\dfrac{y^n}{n!}
\end{align*}
Here \texttt{num\_List}=$\{2m+n,n-m\}$, \texttt{denom\_List}=$\{n\}$, \texttt{X}=$x$ and \texttt{Y}=$y$.\\
\textit{In step 1}, the cancellation of parameters does not give birth to simplifications. Thus, one obtains $C(m,n)= \dfrac{\p{a}{2m+n}\p{b}{n-m}}{\p{c}{n}}$.\\
\textit{In step 2}, on gets the rational functions for $H_5$ as
\begin{align*}
\rho(m,n)=-\frac{m (m-n)}{(2 m+n)^2} \hspace{1cm}\text{and}\hspace{1cm}\sigma(m,n)=\frac{n^2}{(n-m) (2 m+n)}
\end{align*}\\
\textit{In step 3}, the rectangle is evaluated as $\rho(1,0) =\dfrac{1}{4}$ and $\sigma(0,1)=1$.\\
\textit{In step 4}, the conditions are evaluated in terms of $t=m/n$ : $\{t>0\land t<1,t>1\}$.\\
\textit{In step 5}, the cases are evaluated for the above conditions,
\begin{align*}
\Big\{ \Big\{ -\frac{(t-1) t}{(2 t+1)^2},-\frac{1}{(t-1) (2 t+1)}  \Big\},\Big\{ \frac{(t-1) t}{(2 t+1)^2},\frac{1}{(t-1) (2 t+1)} \Big\} \Big\}
\end{align*}\\
\textit{In step 6}, the boundaries are derived. There are two boundaries, one for each cases:
\begin{align*}
\Big\{ \left| x\right|  \left(27 \left| y\right| ^2-36 \left| y\right| +8\right)+16 \left| x\right| ^2-\left| y\right| +1,-\left| x\right|  \left(27 \left| y\right| ^2+36 \left| y\right| +8\right)+16 \left| x\right| ^2+\left| y\right| +1\Big\}
\end{align*}\\
\textit{In step 7}, the degrees are found to be identical and equal to 2. The boundaries are solved for $|y|$ with the constraint $|x|< \dfrac{1}{4}$. From the first boundary one gets
\begin{align*}
\Big\{\left| y\right| =\frac{-(1-12 \left| x\right| )^{3/2}+36 \left| x\right| +1}{54 \left| x\right| } , |y|=\frac{(1-12 \left| x\right| )^{3/2}+36 \left| x\right| +1}{54\left| x\right| } \Big\}
\end{align*}
and from the second
\begin{align*}
\Big\{\frac{(12 \left| x\right| +1)^{3/2}+1}{\left| x\right| }=54 \left| y\right| +36  \Big\}
\end{align*}
Replacing equal signs by the less than signs we find the ROC of $H_5$ as
\begin{align*}
&\left| x\right| <\frac{1}{4}\land \left| y\right| <1\\
&\land \left| y\right| <\left(
\begin{array}{cc}
\Bigg\{ & 
\begin{array}{cc}
\text{Min} \left(\frac{-(1-12 \left| x\right| )^{3/2}+36 \left| x\right| +1}{54 \left| x\right| },\frac{(1-12 \left| x\right| )^{3/2}+36 \left| x\right| +1}{54 \left| x\right| },\frac{(12 \left| x\right| +1)^{3/2}-36 \left| x\right| +1}{54 \left| x\right| }\right) &\text{if}  \left| x\right| <\frac{1}{12} \\
\frac{(12 \left| x\right| +1)^{3/2}-36 \left| x\right| +1}{54 \left| x\right| } & \text{True} \\
\end{array}
\\
\end{array}
\right)
\end{align*}
which can be checked from the code
\begin{Verbatim}[fontsize=\small,frame=single]
In[]:=ROC2[m, n, {2 m + n, n - m}, {n}, x, y]

Out[]={{R,S}, Cartesian Curve, ROC -> ,{1/4,1},
{Abs[y]==(1-(1-12 Abs[x])^(3/2)+36 Abs[x])/...}}
\end{Verbatim}

\subsection{Tests of \texttt{ROC2.wl}}
We now give a comprehensive list of the known results for which the package \texttt{ROC2.wl} was tested. This list includes the 14 Horn's Series $F_1$, ..., $F_4$, $G_1$, $G_2$, $G_3$, $H_1$, ..., $H_7$, seven confluent hypergeometric series and some Kamp\'e de F\'eriet series.

It has to be noted that sometimes the expressions of the ROC obtained form the package can be different from the one given in literature. However, when plotted they have the same shape. For some cases the package gives results in terms of a piecewise function. Those can be written using the Minimum function in order to match with the result from the literature. 

For example the ROC of $H_5$ found in the previous section  can be conveniently written as 

\begin{equation}
r < \frac{1}{4} \land s < 1 \land s< \text{min}\left\lbrack \frac{-(1-12 r)^{3/2}+36 r+1}{54 r}, \frac{(1-12 r)^{3/2}+36 r+1}{54 r},\frac{(12 r+1)^{3/2}-36 r+1}{54 r}\right \rbrack 
\end{equation}

Although both of the forms are essentially the same, the piecewise form is necessary in order to plot the ROC with \textit{Mathematica}.

\section{Conclusions}

We have presented the package \texttt{Olsson.wl}, that can be used to find linear transformations for a large class of multivariable hypergeometric functions, and we have provided a tutorial on its usage. 

Although the obtained \acs can be viewed as functional relations, they are derived from the series representations of the multivariable hypergeometric functions under study. Due to the lack of a good knowledge of the new functions appearing in the linear transformations resulting from this procedure, the latter formulae are valid in only certain ranges of values of the arguments of these functions (the regions of convergence of the corresponding series), which in principle can be found using Horn's theorem, allowing to solve the puzzle of the construction of the starting point function piece after piece. Still, given a multivariable hypergeometric series, finding its region of convergence can be a very difficult task. The companion package \texttt{ROC2.wl} provides the first automatic evaluation of regions of convergence for double hypergeometric series. Constructing algorithms and building a \textit{Mathematica} package for obtaining the regions of convergence of \mhs with more than 2 variables is a work in progress \cite{ASST}. 

The \acs derived from the \texttt{Olsson.wl} package are valid for complex values of the arguments of the involved multivariable hypergeometric series. However, the numerical evaluation of these formulae has to be done with care due to the issues of multivaluedness and branch cuts of the prefactors that appear in them. Therefore, the set of \acs can be combined to built a numerical package for a general hypergeometric function only after resolving these issues. A few numerical packages are available in \textit{Mathematica} and \textit{Maple }for the simplest double hypergeometric functions of the Appell type, and those are far from perfect. While all the four Appell functions are implemented in \textit{Maple}, only the
Appell $F_1$ function is implemented in  the kernel of \textit{Mathematica} (there are also external packages for $F_1$ \cite{Colavecchia2001,Colavecchia2004}). The case of Appell $F_2$ has been recently considered by the present authors in \cite{Ananthanarayan:2021bqz} for real values of its arguments, in non logarithmic situations.  We plan to built a numerical package for all the order two complete double variable hypergeometric functions and Lauricella functions in three variables, valid for complex values of their arguments, in the near future.

\printbibliography

\end{document}